\documentclass[12pt]{iopart}

\usepackage{amssymb,graphicx,epsfig}

\DeclareFontFamily{OMS}{rsfs}{}
\DeclareFontShape{OMS}{rsfs}{m}{n}{%
	<5> rsfs5
	<6> rsfs6
	<7> rsfs7
	<8> rsfs8
	<9> rsfs9
	<10-> rsfs10}{}
\DeclareMathAlphabet{\mathscr}{OMS}{rsfs}{m}{n}
\DeclareMathAlphabet{\mathbss}{OT1}{cmss}{bx}{n}

\def\threej#1#2#3#4#5#6{\left(\matrix{
				 #1&\;#2&\;#3 \cr
				 #4&\;#5&\;#6 }
			    \right)}

\def\sixj#1#2#3#4#5#6{\left\{\matrix{
				 #1&\;#2&\;#3 \cr
				 #4&\;#5&\;#6 }
			    \right\}}

\begin{document}

\title{Scattering polarization of hydrogen lines from electric-induced
atomic alignment}

\author{R. Casini$^1$ and R. Manso Sainz$^2$
\vspace{3pt}}

\address{%
$^1$High Altitude Observatory, National Center for Atmospheric
Research,\footnote{The National Center for Atmospheric Research is
sponsored by the National Science Foundation.}
P.O.~Box 3000, Boulder, CO 80307-3000\break
$^2$Instituto de Astrof\'{\i}sica de Canarias, C/\,V\'{\i}a
L\'actea s/n, E-38205 La Laguna, Tenerife, Spain}


\begin{abstract}
We consider a gas of hydrogen atoms illuminated by a broadband, 
unpolarized radiation with zero anisotropy. In the absence of external
fields, the atomic $J$-levels are thus isotropically populated. While
this condition persists in the presence of a magnetic field, we show 
instead that electric fields can induce the alignment of those levels.
We also show that this \textit{electric alignment} cannot occur in a 
two-term model of hydrogen (e.g., if only the Ly$_\alpha$ transition 
is excited), or if the level populations are distributed according to 
Boltzmann's law. 
\end{abstract}

\pacs{32.70.Jz,32.60.+i,32.30.-r}

\maketitle

\section{Introduction}

Atomic polarization is the property that describes the condition
of population imbalances and quantum coherences between atomic 
levels \cite{Ha72,Bl81}. Population imbalances in a $J$-level 
occur when its Zeeman sublevels, characterized by the azimuthal 
quantum number $M=-J,\ldots,J$, are differently populated. Quantum
coherences (or interferences) occur instead when definite phase 
relations exist between the wave functions representing two distinct 
Zeeman sublevels. An atomic $J$-level is said to be
\emph{isotropically} (or \emph{naturally}) populated if its Zeeman 
sublevels have identical populations, and the quantum coherences 
among them are completely relaxed.

The study of atomic polarization has played a fundamental role in 
the quantum mechanical understanding of the photon-atom interaction 
\cite{Ha24,MZ34,Lo83,MS91,CT92,Ha72}, and in devising techniques 
for controlling and manipulating atoms through radiation (e.g., 
laser cooling \cite{DC89,Mo90}).
Because atomic polarization is affected by the physical
conditions of the atomic gas (e.g., illumination conditions, 
external fields, gas density, collisions), and it manifests 
itself in the polarization of the light emitted by the atoms, its 
study is also fundamental for the diagnostics of plasmas, both in the 
laboratory and in space \cite{FI07,St94,LL04}. 

In order to represent atomic polarization, it is convenient 
to adopt the density-matrix formalism. The atomic density matrix 
projected onto the subspace of a level of total angular momentum $J$ 
is represented by the $(2J+1)^2$ elements $\rho_J(M,M')$, which 
satisfy the conjugation property $\rho_J(M,M')^\ast=\rho_J(M',M)$, 
characterizing Hermitian matrices. The diagonal elements ($M'=M$) 
are proportional to the 
populations of the $M$-states, whereas the off-diagonal elements 
($M'\ne M$) represent the quantum coherences between different 
$M$-states. Hence, for an isotropically populated $J$-level, the 
density matrix is simply proportional to the identity matrix of 
rank $2J+1$.
A $J$-level is further said to be \emph{oriented} if
$\rho_J(-M,-M)\ne\rho_J(M,M)$, and \emph{aligned} if sublevels 
corresponding to different values of $|M|$ have different
populations. In general, an atomic level (and hence the atomic 
system) can be both oriented and aligned. The concept of
orientation and alignment of an atomic level will be further 
clarified in the next section.

Tipically, there are two mechanisms considered in the literature, 
by which atomic polarization can be created:
1) anisotropic and/or polarized excitation by radiation or particles, 
which selectively populates the different magnetic sublevels in the 
transition; 2) in the presence of external fields (electric 
and/or magnetic), a transition is excited by radiation that has 
spectral structure across the frequency range of level splitting
\cite{Ha72,FM73,He85,CT75,Ka50,Ka54,WH97,BB57}. 
In this paper we consider instead a different mechanism, which is the 
alignment of hydrogen levels in the presence of an electric field, when 
the atoms are illuminated with a broadband, unpolarized radiation 
with zero anisotropy
($\oint \frac{d\bi{\Omega}}{4\pi}\,(3\cos^2\vartheta-1)\,
I(\bi{\Omega})\equiv0$, where $I(\bi{\Omega})$ is the radiation 
intensity, and $\bi{\Omega}\equiv(\vartheta,\varphi)$ the propagation 
vector\footnote{A possible realization of such illumination condition 
is that of an atom lying on the surface of a black body.}).
This phenomenon of {\em electric alignment}
has important implications for the diagnostics of electric and
magnetic fields in hydrogen plasmas,
because it affects the scattering polarization that is produced by 
anisotropic irradiation and modified by the presence of external fields.
Interestingly, there is no magnetic counterpart of this mechanism. 
However, if a magnetic field is present simultaneously with the 
electric field, electric alignment can be converted into atomic 
orientation via the alignment-to-orientation (A-O) conversion 
mechanism \cite{Ke84}, resulting in broadband circular polarization 
(BCP) of the scattered radiation.

It is important to remark that electric alignment does not violate 
any conservation or symmetry principle. In particular, it cannot occur
if the atom is illuminated by Planckian radiation, or if the electric 
fields are isotropically distributed (e.g., Holtsmark-type fields 
\cite{Gr74}) and no magnetic field is present. 


\section{The physics of electric alignment.}

We study the mechanism by which an electric field can induce 
atomic aligment in an ensemble of hydrogen atoms that are initially 
in an isotropically populated state because of the particular 
illumination conditions. This process
bears a resemblance with the phenomenon of polarized emission of
Ly$_\alpha$ radiation by electric quenching of hydrogen atoms 
prepared in the 2S$_{1/2}$ metastable state \cite{Fi68,Ha89}, since
both phenomena are determined by the electric mixing of atomic states
of different parity. However, the case studied in this paper has a
much broader impact, because the electric-induced polarized emission 
of the hydrogen lines (not restricted to Ly$_\alpha$) is 
attained in a stationary regime of radiative 
excitation of the hydrogen atoms (rather than as a transient phenomenon 
like in the electric quenching of metastable hydrogen atoms), under 
conditions that are very common in hydrogen plasmas, both in 
the laboratory and in astrophysical objects. In addition, from
our analysis we derive a necessary condition for the generation of 
electric alignment, which applies to any hydrogen line, and 
which to our knowledge has not been clarified in previous work.

Hydrogen atoms in an isotropically populated state are 
completely described by the lowest multipole order ($K=0$) of the 
set of irreducible tensor components of the atomic density matrix 
(also known as \emph{statistical tensors}\footnote{We 
adopt the formalism of the statistical tensors, rather 
than the more common angular-momentum representation introduced 
earlier, because better suited to the description of polarization 
processes. For an extensive review and applications of this formalism, 
see \cite{Bl81,LL04}.}),
\begin{eqnarray}
{}^{nS}\rho^K_Q(LJ,L'J')
&=&\sum_{MM'}(-1)^{J-M}\sqrt{2K+1}
	\threej{J}{J'}{K}{-M}{M'}{Q}\\ \nonumber
&&\hphantom{\sum_{MM'}}
	\times {}^{nS}\rho(LJM,L'J'M')\;,
\end{eqnarray}
where $n$ is the principal quantum number 
of the level of interest, and $L$, $S$, and $J$, are the 
orbital, spin, and total angular momentum quantum numbers, 
respectively.
In fact, the quantity 
$N_{nL}(J)=\sqrt{2J+1}\;{}^{nS}\rho^0_0(LJ,LJ)$ represents the total
population of the level $J$ in the atomic term $nL$. A
direct evalutation of eq.~(1) also shows that our definition of atomic
orientation implies that the $K=1$ multipole of the statistical tensor 
is not vanishing, whereas atomic alignment implies a non-zero $K=2$ 
multipole. Sometimes one speaks more generally of orientation (alignment) 
of an atomic system when at least one of the odd (even) multipoles of the 
statistical tensor is not vanishing \cite{Za88}.

The statistical equilibrium (SE) of hydrogen atoms subject 
to external electric and magnetic fields is governed by the following 
evolution equation for the atomic density operator, $\rho$:
\begin{equation}
\label{eq:evolution}
\dot\rho=({\rm i}\hbar)^{-1}\left[H_0+H_E+H_B,\rho\right]
	-\left(\Gamma\rho+\rho\,\Gamma\right)+{\mathscr T}\rho\;.
\end{equation}
A derivation of eq.~(\ref{eq:evolution}) within the formalism of the
statistical tensors is given in \cite{Ca05}.
In eq.~(\ref{eq:evolution}), $H_0$ is the field-free atomic Hamiltonian,
$H_E=-e_0\bi{r}\cdot\bi{E}$ and $H_B=\mu_0\bi{B}\cdot(\bi{J}+\bi{S})$ 
are the usual electric and magnetic Hamiltonians, and finally
$\Gamma$ and ${\mathscr T}$ are the two radiation operators that are
responsible, respectively, for radiation damping and population 
transfer. (In this work, we neglect the role of
collisions.) In the absence of external fields, and in the presence of
a broadband, unpolarized radiation with zero anisotropy, 
eq.~(\ref{eq:evolution}) reduces to the well-known rate equations for 
the atomic populations. The assumption of broadband illumination
of the atoms implies that the pumping radiation has no spectral structure 
across the frequency range of any given hydrogen line. It is well known 
\cite{He54} that under this hypothesis the scattering process can be 
described as the succession of the two incoherent mechanisms of 
radiation absorption and re-emission, even in the absence of collisions. 
Such \emph{flat-spectrum approximation} is indeed verified in most
astrophysical problems---as well as in many applications to laboratory 
plasmas, except in the case of laser excitation---and it is a
fundamental assumption in the derivation of the results of this work.

\begin{figure}[!t]
\flushright \leavevmode
\includegraphics[width=.55\hsize]{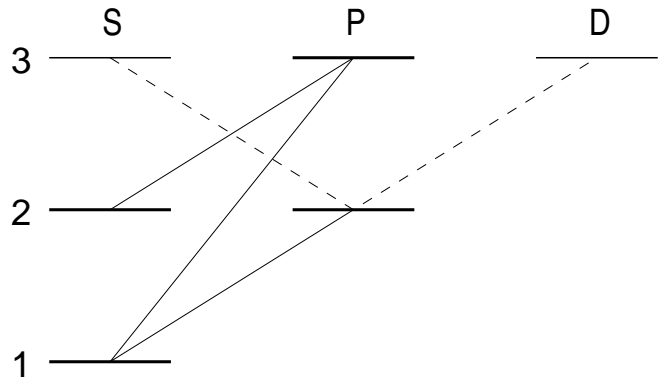}
\caption{\label{fig:model}
Schematic model of the hydrogen atom comprising the first three Bohr levels. 
The transitions indicated with solid lines define the minimal 
set of radiation processes needed to populate the terms 2S and 2P,
in the absence of electric fields. This set
forms the basis for
the model of hydrogen described by the SE system of 
eqs.~(\ref{eq:SE.1})--(\ref{eq:SE.6}).}
\end{figure}

By studying the explicit expressions of the magnetic and electric 
contributions to eq.~(\ref{eq:evolution}) (see eqs.~[17b] and [17c] 
in \cite{Ca05}) one can see how an external field may be able to 
transform atomic populations, ${}^{nS}\rho^0_0(LJ,LJ)$, into quantum 
coherences, under the form of non-diagonal orientation components, 
${}^{nS}\rho^1_Q(LJ,L'J')$. 
In the case of a magnetic field, these orientation components 
can only be generated between $J$ levels that belong to the same $nL$ 
term (because the magnetic Hamiltonian is diagonal with respect to
$L$). On the other hand, since by assumption the incident radiation 
is spectrally flat across the frequency range of any $nL$--$n'L'$ 
transition, the population of these levels always satisfy the condition 
$N_{nL}(J)/N_{nL}(J')=(2J+1)/(2J'+1)$, characteristic of
thermodynamic equilibrium (TE).
In such case, one can show that the total 
magnetic contribution to atomic orientation vanishes. Hence, 
within the flat-spectrum approximation, a magnetic field cannot, 
by itself, generate atomic polarization in an atomic system that is 
initially in an isotropically populated state. This is a 
well-known fact, 
which is at the basis of the application of the Hanle effect to the 
magnetic diagnostics of plasmas \cite{St94,LL04}.

In the presence of an electric field, instead, the electric 
Hamiltonian mixes levels with $\Delta L=1$, and it is found that 
population imbalances (i.e., out of TE)
between such interfering levels can effectively be transformed 
into atomic polarization. 
This process is only inhibited under strict TE conditions for the 
Bohr level $n$, when $N_{nL}/N_{nL'}=(2L+1)/(2L'+1)$.
However, in the absence of collisions, this can only happen if the 
illumination is Planckian ($I(\bi{\Omega})=B_T$; in such case the 
TE distribution of level populations applies to the entire atomic 
model). Otherwise, we must expect a contribution from electric alignment 
to the polarized radiation scattered by a gas of hydrogen atoms, under 
a wide range of physical conditions that are commonly found in 
laboratory and astrophysical plasmas.





In order to quantify this mechanism, we consider a simplified model 
of the hydrogen atom, consisting of the first two Bohr levels, plus the 
3P term (see Fig.~\ref{fig:model}). For simplicity, we neglect the 
fine structure of hydrogen (both spin-orbit interaction and Lamb shift), 
and we also assume that the 3P term is isotropically populated. 
(This last assumption further reduces the dimensionality of the problem.)
The SE of this restricted model of hydrogen---for broadband,
unpolarized incident radiation with zero anisotropy, and in the presence 
of an electric field (which defines the quantization axis)---is governed by 
6 rate equations derived from eq.~(\ref{eq:evolution}). These equations 
involve the populations of the four levels considered in the model, the 
alignment of the 2P level ($a_{\rm 2P}$), and the imaginary 
part\footnote{The real part is zero, if there is no Lamb shift between 
the 2S and 2P levels.} 
of the atomic orientation coherence, ${}^n\rho^1_0(L,L')$, between the 
levels 2S and 2P ($c_{\rm 2S,2P}$).
After imposing the stationary condition $\dot\rho=0$, 
eq.~(\ref{eq:evolution}) yields the following linear system 
\numparts
\begin{eqnarray}
\label{eq:SE.1}
(R_{12}+R_{13})N_{\rm 1S}-R_{21}N_{\rm 2P}-R_{31}N_{\rm 3P}&=&0\;,\\
R_{23}N_{\rm 2S}-R_{32}N_{\rm 3P}-6\omega_E\,c_{\rm 2S,2P}&=&0\;,\\
R_{12}N_{\rm 1S}-R_{21}N_{\rm 2P}-6\omega_E\,c_{\rm 2S,2P}&=&0\;,\\
R_{13}N_{\rm 1S}+R_{23}N_{\rm 2S}-(R_{31}+R_{32})N_{\rm 3P}&=&0\;,\\
R_{21}a_{\rm 2P}-2\sqrt{6}\,\omega_E\,c_{\rm 2S,2P}&=&0\;,\\
\label{eq:SE.6}
3\omega_E N_{\rm 2S}-\omega_E N_{\rm 2P} +\sqrt{6}\,\omega_E\,a_{\rm
2P}+{\textstyle\frac{1}{2}}(R_{23}+R_{21})\,c_{\rm 2S,2P}&=&0\;,
\end{eqnarray}
\endnumparts
where we indicated with $R_{nn'}$ the radiative rate for the transition 
from Bohr's level $n$ to $n'$, and with $\omega_E=a_0 e_0 E/\hbar$ 
the angular frequency associated with the electric field
strength. To completely determine
the solution of this linear system, we must impose the conservation 
of the total atomic population, 
$N_{\rm 1S}+N_{\rm 2S}+N_{\rm 2P}+N_{\rm 3P}=1$. Solving the linear
system algebraically, we find
%
\numparts
\begin{eqnarray}
\label{eq:align_toy}
a_{\rm 2P}&=&\frac{2\sqrt{6}}{R_{21}}\,
c_{\rm 2S,2P}\,\omega_E\;,\\
\label{eq:orien_toy}
c_{\rm 2S,2P}&=&\frac{2 R_{21}\,\omega_E}
	{R_{21}(R_{21}+R_{23})+24\,\omega_E^2} \left(
N_{\rm 2P}-3 N_{\rm 2S}\right)\;.\kern .8cm
\end{eqnarray}
\endnumparts
%
These relations show that the alignment of the 2P term 
is a direct consequence of the quantum coherence between the 2S and 
2P terms due to electric mixing of the same terms, as
expected. However, under TE conditions, electric alignment 
cannot be generated, because $N_{\rm 2P}/N_{\rm 2S}=3$.
In the limit of very strong fields, the quantum coherence between
the 2S and 2P terms vanishes, whereas the atomic alignment of the 2P
term reaches the asymptotic value of
\begin{equation}
\label{eq:align_infinity}
a_{\rm 2P}(\omega_E\to\infty)={\textstyle\frac{1}{\sqrt{6}}} \left(
N_{\rm 2P}-3 N_{\rm 2S} \right)\;,
\end{equation}
which again is zero under TE conditions.

\begin{figure}[!t]
\flushright \leavevmode
\includegraphics[width=.65\hsize]{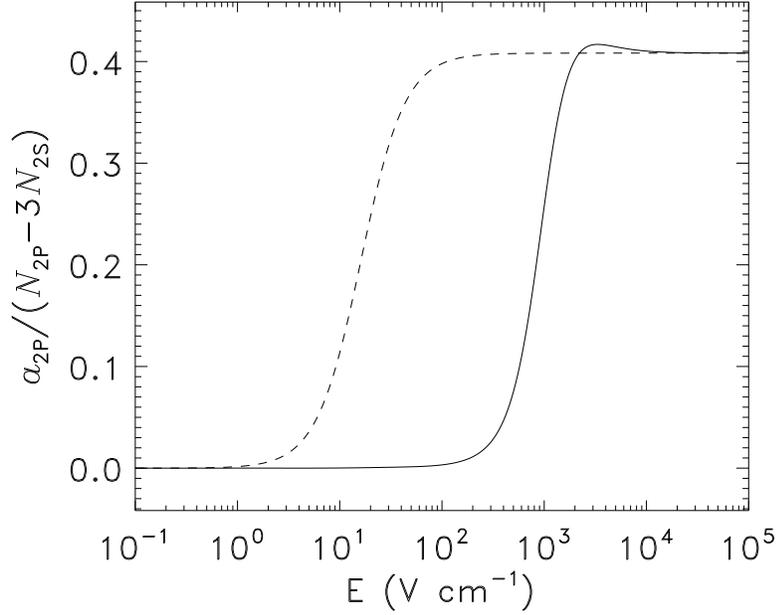}
\caption{\label{fig:align}
Alignment of the 2P term of hydrogen as a function of 
the electric field strength (solid line). The atomic 
model includes all Bohr's levels up to $n=4$ with fine structure. 
The dashed line shows the same quantity for the restricted model of 
eqs.~(\ref{eq:SE.1})--(\ref{eq:SE.6}).}
\end{figure}

For comparison, we calculated numerically the alignment of the 
2P term for a model of hydrogen that includes all Bohr's levels 
up to $n=4$ with fine structure (the dimension of the SE system 
is 1416$\times$1416 in such case). This quantity depends on the
alignment of the fine-structured levels, according to the formula

\begin{eqnarray}
\label{eq:ganza!}
{}^n\rho^K_Q(L,L')&=&\sum_{JJ'} (-1)^{K+L'+J'+S}\sqrt{(2J+1)(2J'+1)} 
	\nonumber \\
\noalign{\vskip -6pt}
&&\hphantom{\sum_{JJ'}}
	\times \sixj{L}{L'}{K}{J'}{J}{S}\,{}^{nS}\rho^K_Q(LJ,L'J')\;.
\end{eqnarray}
which gives in our case
\begin{equation}
\label{eq:align.2P}
a_{\rm 2P}\equiv\rho^2_0(1,1)
	=\sqrt{{\textstyle\frac{2}{3}}}\Bigl[\textstyle
	\rho^2_0\left(1\frac{3}{2},1\frac{3}{2}\right)
	-\rho^2_0\left(1\frac{1}{2},1\frac{3}{2}\right) 
	+\rho^2_0\left(1\frac{3}{2},1\frac{1}{2}\right)\Bigr]
\end{equation}
(for simplicity of notation, we suppressed the configuration superscript).
In Fig.~\ref{fig:align}, we show the alignment 
of the 2P term normalized by the quantity $(N_{\rm 2P}-3 N_{\rm 2S})$, 
as a function of the electric field strength. The illumination conditions 
are such that $I(\bi{\Omega})=(1/2)B_{T=20000\,\rm K}$ with zero anisotropy.
The solid line shows the alignment calculated with
eq.~(\ref{eq:align.2P}), whereas the
dashed line shows the alignment computed according to 
eqs.~(\ref{eq:align_toy}) and (\ref{eq:orien_toy}). We see that 
eq.~(\ref{eq:align_infinity}) gives the correct strong-field limit also 
in the case of a realistic model of hydrogen. The disagreement between 
the two cases for intermediate field strengths is mainly due to 
the omission of the fine structure in the model of 
eqs.~(\ref{eq:SE.1})--(\ref{eq:SE.6}).

%
From eqs.~(\ref{eq:align_toy}) and (\ref{eq:orien_toy}), we can also 
conclude that 
both the alignment of the 2P term and the coherence between the 2S and 
2P terms vanish identically when we restrict the atomic model below the 
minimal set of levels highlighted in 
Fig.~\ref{fig:model}. In fact, 
one can show that the quantity $(N_{\rm 2P}-3 N_{\rm 2S})$ 
contains a factor $(R_{12}R_{23}R_{31}-3R_{13}R_{32}R_{21})$, which is zero 
if we eliminate the 3P term.
It follows that electric alignment cannot be produced in a two-term 
model of hydrogen, for example, when the atoms are illuminated only by 
Ly$_\alpha$ radiation.\footnote{On the other hand, it is sufficient 
to pump the model of eqs.~(\ref{eq:SE.1})--(\ref{eq:SE.6}) with 
just Ly$_\beta$ radiation, to be able to induce electric alignment 
in the 2P term. In that case, we reproduce the results for the
electric quenching of metastable hydrogen atoms previously
considered in the literature \cite{Fi68,Ha89}. Specifically, 
for very large field strengths the Ly$_\alpha$ radiation is found to
be completely polarized along the applied electric field 
($P_{\rm linear}=-100\%$), whereas for vanishing field strengths 
(taking into account the fine structure of hydrogen) we find 
$P_{\rm linear}\approx+32.88\%$, in agreement with the 
results of \cite{Ha89} for the case of
``diabatic'' quenching considered in that work.}


\begin{figure}[!t]
\flushright \leavevmode
\includegraphics[width=.65\hsize]{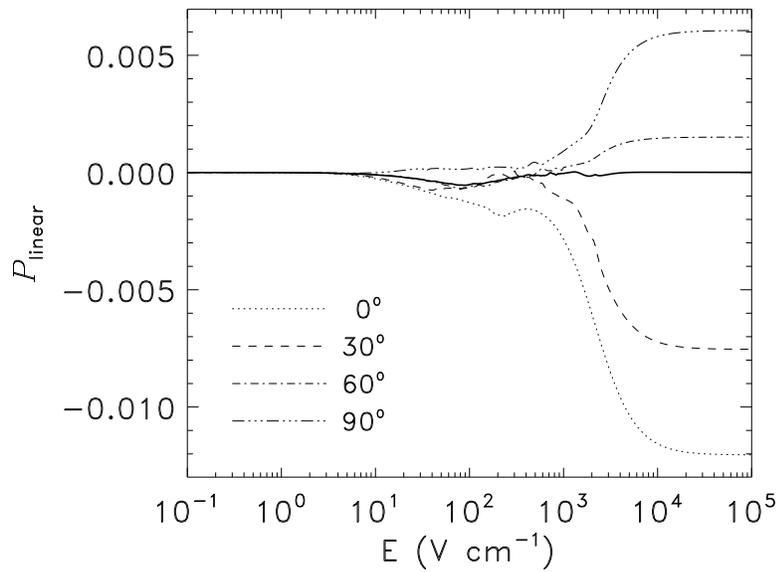}
\caption{\label{fig:E+B=1000.lin}
Degree of BLP from electric alignment of the Ly$_\alpha$ radiation 
scattered at $90^\circ$ from the direction of the quantization axis, 
as a function of the field 
strength of various distributions of electric fields, and in the 
presence of a magnetic field with $B=1000\,\rm G$ directed along 
the quantization axis. The plot shows the case of random-azimuth 
electric fields of different inclinations (dashed curves), and the 
case of isotropic electric fields (thick solid curve).}
\end{figure}

\section{Examples}

We now consider some examples of broadband polarization of 
the scattered radiation of hydrogen lines, resulting from 
electric alignment. All results in this section are calculated
assuming the same illumination conditions and hydrogen model 
(complete up to $n=4$ with fine structure) adopted for 
Fig.~\ref{fig:align}.

Figure~\ref{fig:E+B=1000.lin} shows the broadband linear polarization 
(BLP) of the Ly$_\alpha$ scattered radiation, due to the alignment 
generated by various distribution of electric fields, and in the
presence of a magnetic field with $B=1000\,\rm G$ directed along the
quantization axis. 
The reference direction for positive polarization 
is perpendicular to the quantization axis. 
As a model for the anisotropic distributions of 
electric fields, we considered the case of random-azimuth fields with 
various inclinations from the quantization axis
($\vartheta_E=0^\circ,30^\circ,60^\circ,90^\circ$). For such 
distributions the hydrogen levels are aligned, resulting 
in BLP of the scattered radiation (dashed curves). The alignment 
is zero for vanishing electric fields, 
as expected, because the coherences ${}^{nS}\rho^1_0(LJ,L'J')$ are zero
in that case. On the contrary, the alignment reaches asymptotically
finite values for very large electric fields, in agreement with
eq.~(\ref{eq:align_infinity}). 
In the case of isotropic electric fields (thick solid curve), 
the BLP generally does not vanish, because different field 
directions in the isotropic distribution are not equivalent when 
the spherical symmetry is already broken by the presence of 
the magnetic field. However, it vanishes for very strong electric 
fields, as expected for symmetry reasons.

\begin{figure}[!t]
\flushright \leavevmode
\includegraphics[width=.65\hsize]{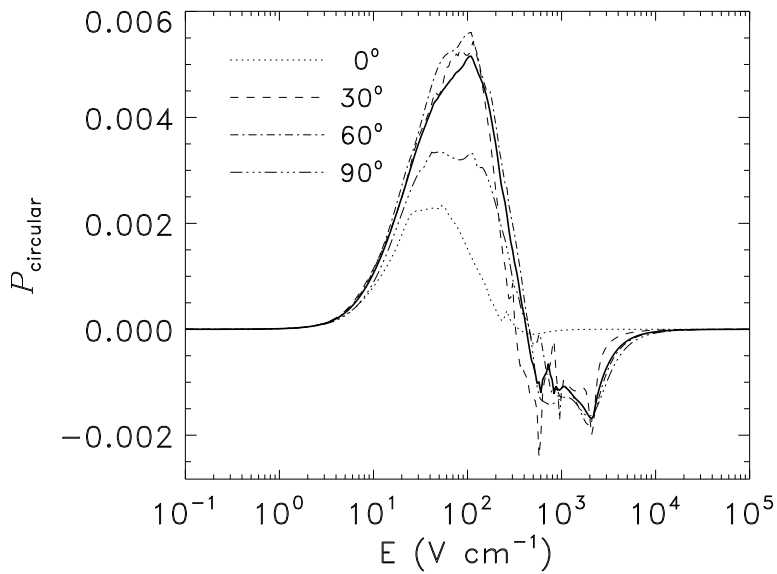}
\caption{\label{fig:E+B=1000.cir}
Same as Fig.~\ref{fig:E+B=1000.lin}, but for the degree of BCP 
observed along the magnetic field.}
\end{figure}

Figure \ref{fig:E+B=1000.cir} shows the BCP of Ly$_\alpha$, for 
the same distributions of the electric and magnetic fields of 
Fig.~\ref{fig:E+B=1000.lin}. We see that the presence of the 
magnetic field is capable of converting the electric alignment of 
the hydrogen levels into atomic orientation via the A-O mechanism
\cite{Ke84}. This orientation is responsible for the 
appearance of BCP in the scattered radiation. On the contrary, 
the electric alignment generated in the presence of only electric 
fields cannot be further converted into BCP-producing atomic 
orientation, so all curves of Fig.~\ref{fig:E+B=1000.cir} would 
collapse to zero in the absence of the magnetic field. 

\section{Conclusions}

In this paper we demonstrated that it is possible to generate
atomic alignment of hydrogen levels \emph{radiatively}
through a mechanism that does not require anisotropic illumination, 
polarization, or spectrally modulated radiation. The alignment 
is instead generated by an electric field that mixes atomic 
levels having imbalanced populations (out of TE).

In practice, such electric alignment will always contribute alongside 
with other competing polarizing mechanisms (for example, anisotropic 
excitation), and the relative importance of this effect will depend on
the physical conditions of the plasma and on the particular hydrogen
line. A recent investigation of the polarization effects of an
isotropic distribution of electric fields in magnetized plasmas,
where all these competing mechanisms are taken into account, has 
been presented in \cite{CM06} for the case of the Ly$_\alpha$ and
H$_\alpha$ lines.




\section*{References}

\begin{thebibliography}{24}

\bibitem{Ha72}
Happer W 1972 \RMP \textbf{44} 169
\bibitem{Bl81}
Blum K 1981 \textit{Density Matrix Theory and Applications} (New York:
Plenum Press)

\bibitem{Ha24}
Hanle W 1924 \textit{Z.\ Phys.} \textbf{30} 93 \bibitem{MZ34}
Mitchell A C G and Zemansky M W 1934 \textit{Resonance Radiation
and Excited Atoms} (Cambridge: Cambridge Press)
\bibitem{Lo83}
Loudon R 1983 \textit{The Quantum Theory of Light (2nd ed.)} (Oxford:
Clarendon Press)
\bibitem{MS91}
G.\ Moruzzi G and Strumia F 1991 \textit{The Hanle Effect and Level-Crossing
Spectroscopy} (New York: Plenum)
\bibitem{CT92}
Cohen-Tannoudji C, Dupont-Roc J and Grynberg G 1992 \textit{Atom-Photon
Interactions} (New York: Wiley)

\bibitem{DC89}
Dalibard J and Cohen-Tannoudji C 1989 \JOSA B \textbf{6} 2023
\bibitem{Mo90}
Monroe C, Swann W, Robinson H and Wieman C 1990 \PRL \textbf{65} 1571

\bibitem{FI07}
Fujimoto T and Iwamae A 2007 \textit{Plasma Polarization Spectroscopy}
(Berlin: Springer)

\bibitem{St94}
Stenflo J O 1994 \textit{Solar Magnetic Fields} (Dordrecht: Kluwer)
\bibitem{LL04}
Landi Degl'Innocenti E and Landolfi M 2004 \textit{Polarization in
Spectral Lines} (Dordrecht: Kluwer)

\bibitem{FM73}
Fano U and Macek J H 1973 \RMP \textbf{45} 553
\bibitem{He85}
Hertel I V, Schmidt H, B\"ahring A and Meyer E 1985  
	\textit{Rep.\ Prog.\ Phys.} \textbf{48} 375
\bibitem{CT75}
Cohen-Tannoudji C 1975 \textit{Atomic Physics 4} ed G zu Putlitz et al 
	(New York: Plenum)
\bibitem{Ka50}
Kastler A 1950 \textit{J.\ Phys.\ Radium} \textbf{11} 255
\bibitem{Ka54}
Kastler A 1954 \PRS A (London) \textbf{67} 853
\bibitem{WH97}
Walker T G and Happer W 1997 \RMP \textbf{69} 629
\bibitem{BB57}
Bell W E and Bloom A L 1957 \PR \textbf{107} 1559

\bibitem{Ke84}
Kemp J C, Macek J H and Nehring F W 1984 \textit{Astrophys.\ J.} 
\textbf{278} 863

\bibitem{Gr74}
Griem H R 1974 \textit{Spectral Line Broadening by Plasmas} (New York:
Academic Press)

\bibitem{Fi68}
Fite W L, Kauppila W E and Ott W R 1968 \PRL \textbf{20} 409
\bibitem{Ha89}
Harbich W, Hippler R, Kleinpoppen H and Lutz H O 1989 \PR~A
\textbf{39} 3388

\bibitem{Za88}
Zare R N 1988 \textit{Angular Momentum} (New York: Wiley)

\bibitem{Ca05}
Casini R 2005 \PR A \textbf{71} 062505

\bibitem{He54}
Heitler W 1954 \textit{The Quantum Theory of Radiation (3rd ed.)}
(Oxford: Clarendon Press)

\bibitem{CM06}
Casini R and Manso Sainz R 2006 \jpb \textbf{39} 3241













\end{thebibliography}
\end{document}